# Accelerating first-principles estimation of thermal conductivity by machine-learning interatomic potentials: A MTP/ShengBTE solution

Bohayra Mortazavi*[a,b], Evgeny V. Podryabinkin[c], Ivan S. Novikov[c,d],

Timon Rabczuk[e], Xiaoying Zhuang**[e,f] and Alexander V. Shapeev[c]

[a]*Chair of Computational Science and Simulation Technology, Department of Mathematics and Physics, Leibniz Universität Hannover, Appelstraße 11,30157 Hannover, Germany.*
[b]*Cluster of Excellence PhoenixD (Photonics, Optics, and Engineering–Innovation Across Disciplines), Gottfried Wilhelm Leibniz Universität Hannover, Hannover, Germany.*
[c]*Skolkovo Institute of Science and Technology, Skolkovo Innovation Center, Nobel St. 3, Moscow 143026, Russia.*
[d]*Institute of Materials Science, University of Stuttgart, Pfaffenwaldring 55, 70569 Stuttgart, Germany.*
[e]*Division of Computational Mechanics, Ton Duc Thang University, Ho Chi Minh City, Vietnam.*
[f]*Faculty of Civil Engineering, Ton Duc Thang University, Ho Chi Minh City, Vietnam.*

**Abstract**

Accurate evaluation of the thermal conductivity of a material can be a challenging task from both experimental and theoretical points of view. In particular for the nanostructured materials, the experimental measurement of thermal conductivity is associated with diverse sources of uncertainty. As a viable alternative to experiment, the combination of density functional theory (DFT) simulations and the solution of Boltzmann transport equation is currently considered as the most trusted approach to examine thermal conductivity. The main bottleneck of the aforementioned method is to acquire the anharmonic interatomic force constants using the computationally demanding DFT calculations. In this work we propose a substantially accelerated approach for the evaluation of anharmonic interatomic force constants via employing machine-learning interatomic potentials (MLIPs) trained over short ab-initio molecular dynamics trajectories. The remarkable accuracy of the proposed accelerated method is confirmed by comparing the estimated thermal conductivities of several bulk and two-dimensional materials with those computed by the full-DFT approach. The MLIP-based method proposed in this study can be employed as a standard tool, which would substantially accelerate and facilitate the estimation of lattice thermal conductivity in comparison with the commonly used full-DFT solution.

Keywords: Machine-learning; Interatomic potentials; Thermal conductivity; First-principles; 2D materials;
Corresponding authors: *bohayra.mortazavi@gmail.com; ** xiaoying.zhuang@tdtu.edu.vn

## 1. Introduction

Thermal conductivity is an essential property of a material, which can play a pivotal role in the engineering design of a product. Nowadays, appropriate thermal management is a common challenge for rapidly-growing fields, like nanoelectronics and electric vehicles. In principle, for the majority of applications, materials with higher thermal conductivities are more desirable in order to facilitate the excessive heat dissipation and avoid overheating issues. For some specific applications like thermal insulating and thermoelectric materials, materials with lower thermal conductivities are nonetheless more favorable to reduce the thermal energy losses and improve thermoelectric figure of merit, respectively. In the engineering design of a product, to accurately examine the temperature evolution during the operation, the thermal conductivity of each building block ought to be known. Therefore, accurate estimation of the thermal conductivity is required to achieve the efficient design and avoid dangerous scenarios stemming from overheating.

During the last decade, two-dimensional (2D) materials have gained remarkable attention as the novel class of materials to improve the design and efficiency of diverse advanced systems. Graphene [1,2], a prototypical member of the 2D materials class, is known to show ultrahigh thermal conductivity [3,4] outperforming all known materials. The exceptionally high thermal conductivity offers the graphene as an excellent candidate to enhance the thermal management in diverse systems, like nanoelectronics and Li-ion batteries[5–8]. Despite high-quality and symmetrical atomic lattice of graphene and after extensive studies accomplished during the last decade, the exact value of its thermal conductivity still remains debated from both experimental and theoretical points of view [9,10]. For the single-layer graphene, the experimentally measured thermal conductivities mostly fall within 1500–5300 W/mK [3,11–13]. In fact, there exist numerous sources of uncertainty in the experimental measurements of the thermal conductivity of 2D materials, usually resulting in the remarkable scattering of reported values. Taking into account the clean and high quality of graphene crystals, it is clear that for more complicated 2D lattices, the uncertainties in the experimentally measured values can be more pronounced. As an alternative to experimental characterizations, the development of an accurate modeling approach is critical for the evaluation of the thermal conductivity of 2D materials.

To the best of our knowledge, the combination of density functional theory (DFT) simulations and the solution of Boltzmann transport equation (BTE) is currently the most trusted approach to evaluate the thermal conductivity of bulk and 2D materials. The BTE solution of thermal conductivity normally requires the evaluation of second- and third-order (anharmonic) interatomic force constants, which are commonly acquired by DFT calculations over supercell lattices. The most computationally demanding step in this approach is to obtain the anharmonic force constants, which depending on the cutoff distance and lattice symmetry may require a few hundred/thousand single-point DFT force calculations. In this study, we propose the employment of machine-learning interatomic potentials (MLIP) to substitute the computationally demanding DFT calculations in obtaining the anharmonic force constants. Due to negligible computational costs of force constant calculations using MLIP, the proposed approach therefore offers a substantial acceleration in the evaluation of anharmonic interatomic force constants in comparison with the DFT-based solution. We used moment tensor potentials (MTPs)[14], as an accurate and computationally efficient model of MLIPs[15–17] in the evaluation of interatomic forces. To the best of our knowledge, the ShengBTE[18] package is currently the most employed package for the full iterative solution of the BTE. In this paper, we present the integrated MTP/ShengBTE method for the practical calculation of the lattice thermal conductivity. The proposed approach was tested on several bulk and 2D materials and shows remarkably close agreement with the results available in the literature on the basis of full-DFT calculations.

## 2. Computational methods

In this work, first-principles DFT calculations were conducted to obtain the phonon dispersion relations, second-order interatomic force constants, and also to create the training sets for the MTPs. To this aim, *Vienna Ab-initio Simulation Package* (VASP)[19–21] and generalized gradient approximation (GGA) with either Perdew–Burke–Ernzerhof (PBE)[22], Perdew-Burke-Ernzerhof revised for solids (PBEsol) [23] or revised Perdew-Burke-Ernzerhof (revPBE) [24] were adopted in the calculations. The geometry optimized structures were acquired using the conjugate gradient method with the convergence criteria of $10^{-5}$ eV and 0.001 eV/Å for the energy and forces, respectively. The PHONOPY code[25] was employed to create the optimal sets of atomic positions for DFT or MTP force calculations and subsequently obtain phonon dispersion relations and second-order interatomic force constants with DFT and MTP-based results as inputs. The interatomic force constants were evaluated by considering the supercell structures. The details of MTP and PHONOPY interface can be found in our latest study [26]. Ab-initio molecular dynamics (AIMD) simulations were performed with a time step of 1 fs. We employed Monkhorst-Pack[27] k-point grids of 3×3×1 and 2×2×2, for 2D and bulk lattices, respectively. The supercell sizes for the phonon and AIMD calculations and corresponding plane-wave cutoff energies will be explicitly mentioned for every considered example.

MTP is a local potential in the sense that the total energy $E$ of an atomistic sample with $N$ atoms is the summation of contributions $V$ of neighborhoods $u_i$ of each $i$-th atom: $E \equiv E^{MTP} = \sum_{i=1}^{N} V(u_i)$. The neighborhood of a central atom is defined as a collection, via:

$$u_i = \left( \{r_{i1}, z_i, z_1\} \dots , \{r_{ij}, z_i, z_j\} \dots , \{r_{iN_{\text{neigh}}}, z_i, z_{N_{\text{neigh}}}\} \right), \quad (1)$$

here the $j$-th atom is referred as a neighbor of the $i$-th (central) atom within the preset cutoff radius $R_{\text{cut}}$, $z_i$ and $z_j$ are the types of the central and neighboring atoms, respectively, $r_{ij}$ is the corresponding interatomic vector and $N_{\text{neigh}}$ is the number of atoms in the neighborhood. The contribution of each central atom and its associated neighborhood to the system's total energy shows: $V(u_i) = \sum_\alpha \xi_\alpha B_\alpha(u_i)$, where $B_\alpha$ are the basis functions and $\xi_\alpha$ are the parameters of a MLIP. Basis functions are constructed according to the all possible contractions of the moment tensor descriptors[14], yielding a scalar as follows:

$$M_{\mu,\nu}(r_i) = \sum_{j=1}^{N_{\text{nei}}} f_\mu \left( |r_{ij}|, z_i, z_j \right) r_{ij}^{\otimes \nu} \quad (2)$$

here, the first factor $f_\mu(|r_{ij}|, z_i, z_j)$ is the radial part of potential MLIP, which only depends on the distance between atoms $i$ and $j$ and their atomic types and "$\otimes$" is the outer product. The radial part is expanded by a set of radial basis functions $\varphi_\beta(|r_{ij}|)$ multiplied by the $(R_{cut-}|r_{ij}|)^2$ smoothing factor.

$$f_\mu(|r_{ij}|, z_i, z_j) = \sum_\beta c^{(\beta)}_{\mu, z_i, z_j} \varphi_\beta(|r_{ij}|)(R_{cut-}|r_{ij}|)^2, \quad (3)$$

here $c^{(\beta)}_{\mu, z_i, z_j}$ are the radial coefficients (parameters). The MTP parameters of $\xi_\alpha$ and $c^{(\beta)}_{\mu, z_i, z_j}$, are acquired by solving the minimization problem of:

$$\sum_{k=1}^{K} \left[ w_e \left( E_k^{AIMD} - E_k^{MTP} \right)^2 + w_f \sum_i^N |f_{k,i}^{AIMD} - f_{k,i}^{MTP}|^2 + w_s \sum_{i,j=1}^{3} |\sigma_{k,ij}^{AIMD} - \sigma_{k,ij}^{MTP}|^2 \right] \rightarrow \min, \quad (4)$$

where $E_k^{AIMD}$, $f_{k,i}^{AIMD}$ and $\sigma_{k,ij}^{AIMD}$ are the systems total energy, atomic forces and atomic stresses in the training set, respectively, with total $K$ number of configurations, $E_k^{MTP}$, $f_{k,i}^{MTP}$ and $\sigma_{k,ij}^{MTP}$ are the corresponding values predicted by the MTP, and $w_e$, $w_f$ and $w_s$ are the optimization problem importance weights for the energies, forces and stresses, respectively, which are set to 1, 0.1 and 0.001, respectively. In this work, interatomic force constants are acquired using the MTPs[14], trained over AIMD runs at different temperatures of 50, 300, 500 and 700 K, each with less than 1000 time steps. The training sets were sampled from the full trajectories (one configuration was sampled every five time steps).

Finally, the lattice thermal conductivity is estimated using the full iterative solutions of the Boltzmann transport equation, as implemented in the ShengBTE [18] package. Harmonic force constants are obtained using the PHONOPY code [25] with DFT results as inputs. Anharmonic interatomic force constants are calculated using the trained MTPs. Since force evaluations with MTP have a negligible cost, we take the same (large) supercell samples for the harmonic and anharmonic force constant calculations. Isotope scattering is considered in all examples, which is essential for direct comparison with experimental data for naturally occurring samples. Born effective charges and dielectric constants contributions in the dynamical matrix are taken into account only for the bulk InAs structure. The convergence of the lattice thermal conductivity with respect to the $q$-mesh grid is also examined. To facilitate the practical application, in the data availability section, the full computational details with developed python scripts are included. There we provide the full set of the input files for all considered examples along with the numerical procedure to extract the anharmonic force constants using the trained MTPs.

## 3. Results and discussions

The main objective of the present study is to examine the accuracy of estimated lattice thermal conductivities with the MTP-based solution in obtaining anharmonic interatomic force constants. In our latest study [26], obtained results for diverse 2D lattices confirm that the MTP-based solution can accurately reproduce the phononic properties in close agreement with density functional perturbation theory (DFPT) simulations. For the validation of the proposed approach, we consider several 2D and bulk lattices and compare MTP-based results with available data in the literature. We first examine the lattice thermal conductivity of graphene, which has been extensively studied experimentally and theoretically during the last decade. By having a glance at the available data in the literature, it becomes conspicuous that despite a rather simple bonding nature, highly symmetrical lattice and high quality of graphene crystals, the exact value of its thermal conductivity still remains debated from both experimental and theoretical points of view. According to full-DFT BTE based studies [28,29], it has been suggested that the type of exchange-correlation functional can yield substantial effects on the estimated thermal conductivity. In the work by Qin *et al.* [28], single-layer graphene's thermal conductivity at room temperature was predicted from 1936 W/mK (with PBE) to 4376 W/mK (with VdW-DF2). In the recent work by Taheri and co-workers [29], they predicted distinctly higher thermal conductivities ranging from 5442 W/mK (with LDA) to 8677 W/mK (with PBEsol). To check the effects of exchange-correlation functional for the case of graphene, we thus consider PBE, PBEsol, and revPBE functionals. In our calculations, we consider a plane-wave cutoff energy of 500 eV and the lattice constant of graphene is found to be 2.467, 2.461, and 2.479 Å, respectively with PBE, PBEsol, and revPBE functionals.

In our interatomic force constants calculations, we consider a 10×10×1 supercell, which includes 200 carbon atoms. The AIMD calculations for the preparation of training sets are conducted over 6×6×1 supercells. As discussed in our recent study[26], incorporation of larger supercells in the AIMD calculations can however secure higher accuracy. MTPs with 901 parameters are passively trained and subsequently used to reproduce the phonon dispersion relations of graphene. In Fig. 1 the phonon dispersion relations of single-layer graphene by MTP and DFPT results are compared, which reveal close agreement for the all three considered functional.

We next examine the lattice thermal conductivity using the trained MTPs to acquire anharmonic interatomic force constants, in which the interactions with elevenths nearest neighbors are considered. The thermal conductivity of graphene at the room temperature is predicted to be 3730, 3640, and 3600 W/mK, respectively, on the basis of trained MTPs with PBE, PBEsol, and revPBE functionals. It is noticeable that unlike the earlier reports [28,29], the effect of exchange-correlation functional on the estimated thermal conductivity value is negligible. In the work by Fugallo *et al.* [30] they calculated the graphene's thermal conductivity to be 3600 W/mK, which is very close to our estimated values. Using the PBE functional and ShengBTE package for the BTE solution, the room temperature thermal conductivity of naturally occurring graphene has been reported to be 1936[28], 3100[31], 3550[32], 3845[33], 3590 [34], 3720[35] and 3288[36], which are generally consistent with our results. One of the main sources of scatterings in different reports is related to the supercell size effect. We found that for the case of graphene the thermal conductivity converges for the supercell size of 8×8×1. In Fig. 1d the MTP-based results for the temperature-dependent thermal conductivity of single-layer graphene are compared with experimental and full-DFT based estimations, which show good agreement. Furthermore, using the MTP-based solution, the temperature power factor for the thermal conductivity of graphene is found to be 1.35, compatible with full-DFT based reports of 1.32 by Lindsay *et al.* [36] and 1.34 by Fugallo *et al.* [30]. Fig. 1e compares the cumulative lattice thermal conductivity of single-layer graphene at the room temperature as a function of mean free path by MTP and full-DFT calculations, which also reveal close trends. For the single-layer graphene, acoustic phonons are known as the dominant heat carriers. In Fig. 1f the contribution of ZA, TA and LA acoustic modes on the total lattice thermal conductivity of graphene by MTP and full-DFT calculations are compared, which also reveal close agreement. It is noticeable once again that MTP-based solutions show negligible effects for different functionals. Despite the remarkable scattering between different experimental and theoretical results for the case of graphene, the conducted comparison clearly reveals the high accuracy of the proposed MTP-based solution.

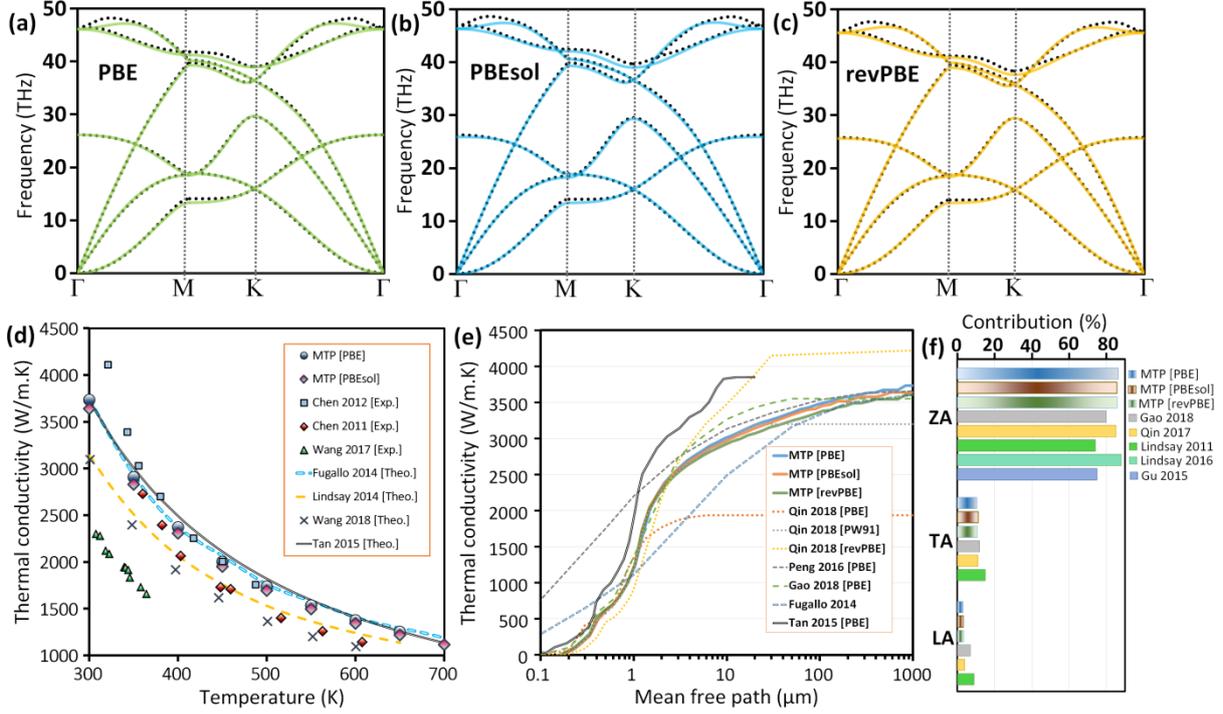

Fig. 1, (a-c) Phonon dispersion relations of graphene with PBE, PBEsol, and revPBE exchange-correlation functionals, respectively, by DFPT (dotted lines) and MTP (continuous lines) methods. (d) Temperature-dependent thermal conductivity of graphene by the MTP-based solutions compared with experimental (Exp.) works by Chen 2011[37], Chen 2012[38], Wang 2017 [39], and theoretical (Theo.) results by Fugallo 2014[30], Lindsay 2014 [36], Wang 2018 [31] and Tan 2015 [33]. The experimental data include error-bars that are not plotted for the better clarity (e) Cumulative lattice thermal conductivity of graphene at the room temperature as a function of mean free path by MTP and full-DFT solutions by Qin 2018 [28], Fugallo 2014 [30], Peng 2016 [35], Gao 2018 [32] and Tan 2015 [33] with different exchange correlation functions. Contribution of different acoustic modes on the total room-temperature thermal conductivity of graphene by MTP and full-DFT studies by Lindsay 2011 [40], Gao 2018 [32], Qin 2017 [41] and Gu 2015[42].

Because of the remarkable scattering in the experimentally measured and theoretically predicted thermal conductivity of graphene and other 2D materials, we next examine the accuracy of the proposed approach for the bulk materials. In this case, we consider silicon and InAs, the examples also considered in the original ShengBTE manuscript by Li *et al.* [18]. We also consider bulk diamond and BAs for further examination. For the silicon, InAs, diamond and BAs plane-wave cutoff energies of 330, 300, 500, and 400 eV are employed, respectively. The lattice constants of silicon and InAs with PBE functional are found to be 5.47 and 6.06 Å, respectively, match closely with those reported by Li *et al.* [18]. For the diamond and BAs, we use the PBEsol functional because the obtained lattice constant of 3.572 and 4.779 Å, respectively, are in close agreement with corresponding experimentally measured values of 3.567 and 4.777[43] Å. We remind that all considered bulk structures show Zinc Blende diamond-like lattices. In line with the work by Li *et al.* [18], harmonic interatomic force constants are calculated using the 5×5×5 supercell samples. AIMD calculations are conducted over rectangular supercells with 144 atoms to create the training sets. MTPs with 901 and 1009 parameters are trained for monoelemental and binary lattices, respectively. In Fig. 2, the predicted phonon dispersion relations for considered bulk lattices by passively fitted MTPs are compared with DFPT-based results, which reveal remarkably close agreement.

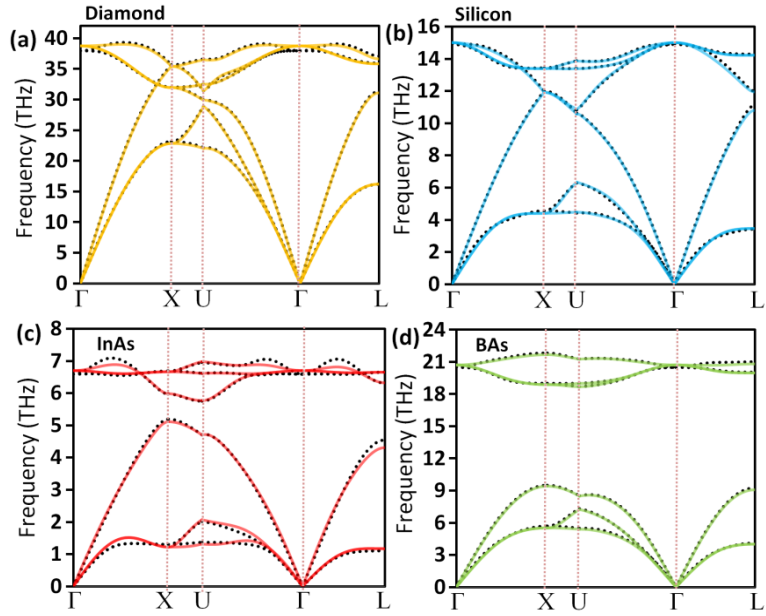

**Fig. 2**, Phonon dispersion relations of bulk diamond, silicon, InAs and BAs obtained using the DFPT (dotted lines) and first-attempt MTPs (continuous lines) methods.

In Fig. 3, the estimated phononic thermal conductivity of the considered bulk structures by the accelerated method are compared with experimental and full-DFT counterparts. The negligible computational costs of the force evaluations with MTP enables using relatively large supercells. Our calculations are done for 5×5×5 supercell structures for the anharmonic force constants calculations, whereas in work by Li *et al.* [18] smaller supercells of 4×4×4 were used. For the consistency with the work by Li *et al.* [18], interactions with the fourth nearest neighbors are considered in these calculations. Worthwhile to remind that dielectric tensor and Born effective charges contributions in the BTE solution are only considered for InAs, with the corresponding values taken from the ShengBTE [18] examples. As it is clear, the results by the proposed accelerated approach are in excellent agreement with experimental and full-DFT based theoretical results. The maximum discrepancy occurs for the case of BAs, around 12%, which can be partially associated due to computational details. For example, in the study by Protik et al. [44], they predicted a room temperature thermal conductivity of 2300 W/mK for naturally occurring BAs, which is only by around 9% difference with our predicted value.

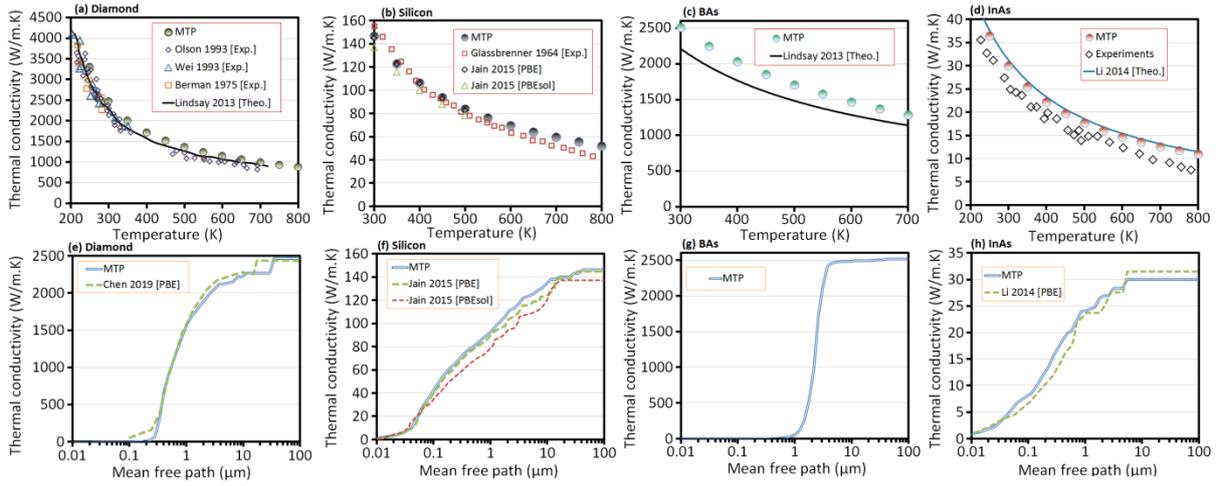

Fig. 3, Thermal conductivity of bulk diamond, silicon, InAs, and BAs by the MTP-based solutions compared with experimental and full-DFT counterparts. Cumulative lattice thermal conductivity as a function of the mean free path is compared in the second row, respectively. The data for the diamond are taken from Olson 1993 [45], Wei 1993 [46], Berman 1975 [47], Lindsay 2013 [48], and Chen 2019 [49]. For the silicon, the data are taken from Glassbrenner 1964 [50] and Jain 2015 [51]. For the BAs the theoretical data are from Lindsay 2013 [48]. The theoretical and experimental data [52,53] for the InAs are taken from the original ShengBTE paper by Li 2014 [18].

Currently, 2D materials are one of the fastest-growing class of materials. The complexity and low-symmetry of novel 2D materials lattices demand for excessive computational costs for the evaluation of anharmonic force constants. In fact, the introduced alternative by this study is highly promising for the assessment of 2D materials thermal conductivity. From the modeling points of view, to deal with computational difficulties in the valuation of anharmonic force constants using the conventional DFT calculations, researchers may consider smaller supercells, interactions with fewer neighbors, coarser K-point grids and/or lower plane-wave cutoff energies. These simplifications can justify the remarkable scattering in the predicted values for the 2D materials' thermal conductivity. Next we examine the results for the thermal conductivity of several 2D materials by the MTP-based accelerated approach, taking into account that the exact values of thermal conductivities are debated. Here we consider single-layer penta-graphene, silicene, phosphorene, F-diamane, and $MoS_2$, as illustrated in Fig. 4. Plane-wave cutoff energies of 500, 330, 400, 500, and 400 eV are considered for penta-graphene, silicene, phosphorene, F-diamane, and $MoS_2$, respectively. All the considered 2D lattices are isotropic, except phosphorene (find Fig. 4c), which shows anisotropic lattice constants of 4.624 and 3.299 Å along the armchair and zigzag directions, respectively. The lattice constants of penta-graphene, silicene, F-diamane, and $MoS_2$ are found to be 3.642, 3.868, 2.546, and 3.184 Å, respectively, all consistent with the reported values in the literature. Worthy to note that F-diamane was most recently experimentally realized by Bakharev and coworkers[54].

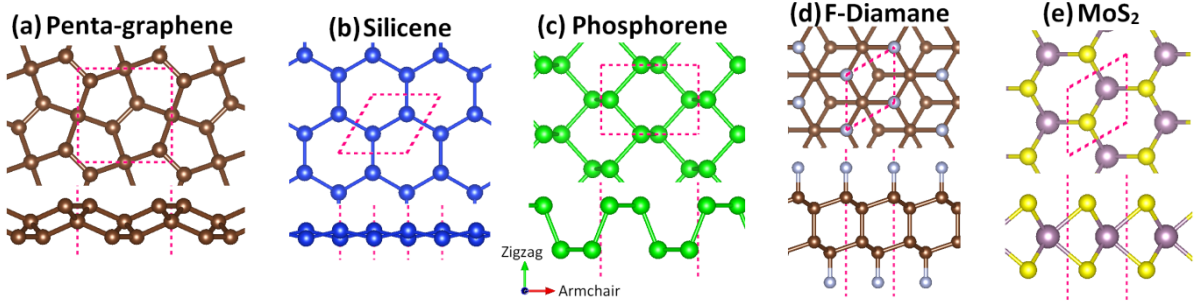

Fig. 4, Top and side views of the considered 2D lattices. Red-dashed lines illustrate the primitive unit cell.

For the preparation of training sets, AIMD simulations are conducted using 4×4×1, 6×6×1, 3×5×1, 4×4×1, and 5×5×1 supercells for penta-graphene, silicene, phosphorene, F-diamane, and MoS$_2$, respectively. Similarly to the bulk structures, MTPs with 901 and 1009 parameters are trained for monoelemental and binary 2D lattices, respectively. For the evaluation of interatomic force constants, 4×4×1, 10×10×1, 4×6×1, 5×5×1 and 6×6×1 supercells are employed for penta-graphene, silicene, phosphorene, F-diamane, and MoS$_2$ monolayers, respectively. The predicted phonon dispersion relations for the considered 2D lattices by the passively fitted MTPs are compared with those by the DFPT method in Fig. 5, which reveal excellent agreement and highlight the high accuracy of MTPs in describing the interatomic forces.

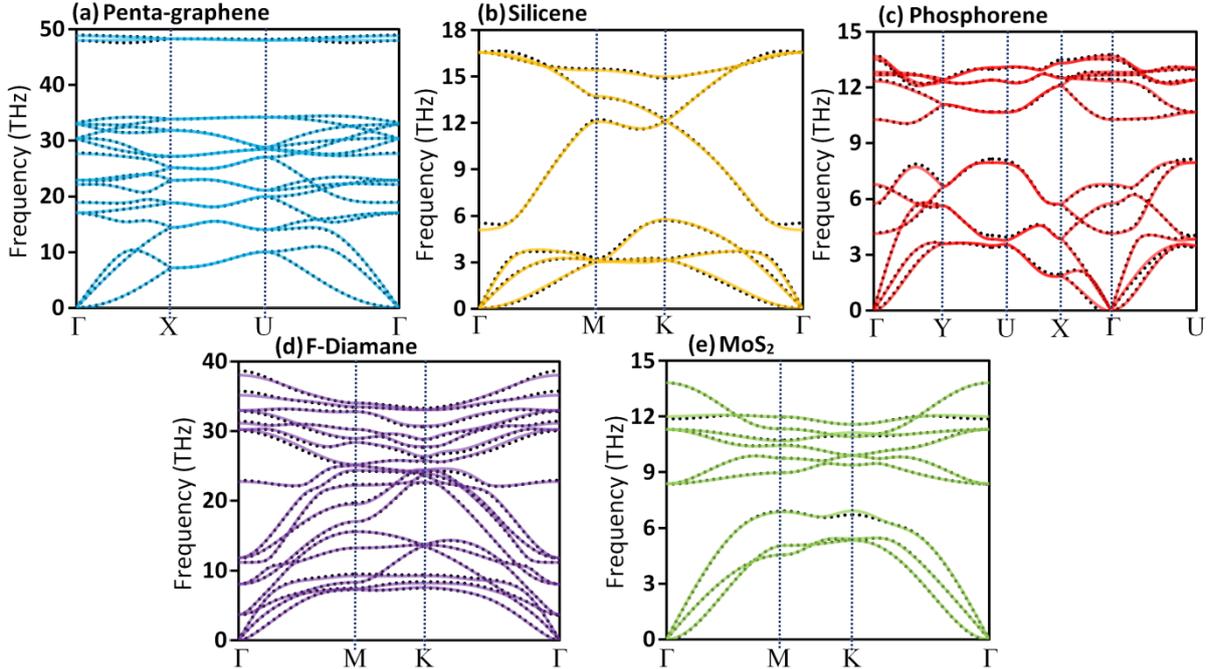

Fig. 5, Phonon dispersion relations of penta-graphene, silicene, phosphorene, F-diamane, and MoS$_2$ acquired by the DFPT (dotted lines) and first-attempt MTPs (continuous lines).

We next examine the lattice thermal conductivity of the considered 2D lattices. Anharmonic interatomic force constants are acquired using the trained MTPs, in which the interactions with the twelfth, eleventh, twelfth, ninth, and eights nearest neighbors are considered for penta-graphene, silicene, phosphorene, F-diamane, and MoS$_2$ monolayers, respectively. In Fig. 6, predicted lattice thermal conductivities of the considered 2D lattices by the MTP-based BTE solution are compared with full-DFT counterparts. Similar to the case of graphene, a remarkable scattering between different DFT-based BTE solutions causes ambiguity in finding the exact values of lattice thermal conductivity. With decreasing of the lattice symmetry, the complexity of the DFT-based solution increases. Therefore to decrease the computational costs, various simplifications are applied in different studies, resulting in scatting of reported values. Let's consider the $MoS_2$, for which the in-plane lattice thermal conductivity of bulk structure is experimentally measured to be 85–110 W/mK [55]. For 2D materials, because of absence of van der Waals phonon scattering from the interactions with adjacent layers[4], the thermal conductivity of monolayers is usually expected to be higher than those of bulk and few-layer structures. This way, the thermal conductivity of single-layer $MoS_2$ is expected to be higher reported experimental range of 85–110 W/mK [55] for the bulk structure, as also theoretically confirmed by Gandi et al. [56] and Gu et al. [57]. However, due to computational setups, DFT-based reports may not satisfy this expectation. For the single-layer $MoS_2$, our proposed approach yields a thermal conductivity of 152 W/mK, very close to the predicted value of ~155 W/mK by Peng et al. [58] and 138 W/mK by Gu et al. [57]. In the study by Gu et al. [57], they reported the value for the sample with a length of 10 μm, which can be probably lower than the converged diffusive thermal conductivity. Comparisons illustrated in Fig. 6 clearly show that the estimated thermal conductivities for the considered monolayers using the MTP-based accelerated method are within the values acquired using the full-DFT based solutions, which confirm the remarkable accuracy of the proposed approach.

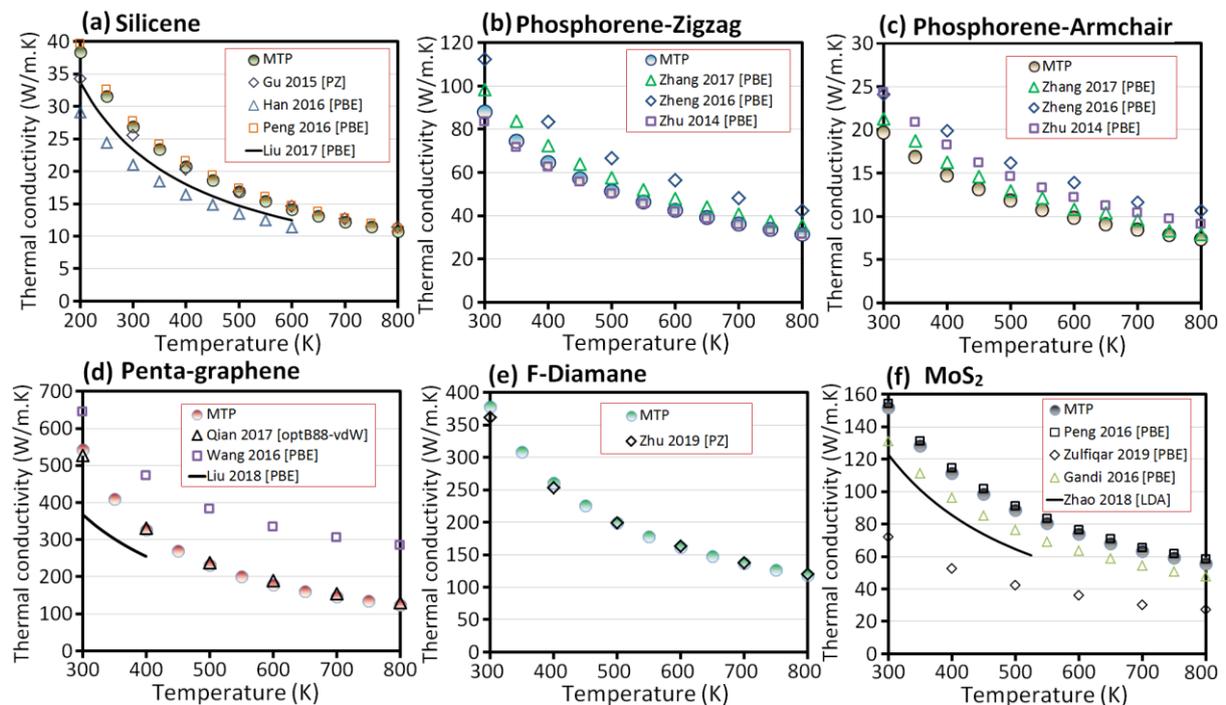

**Fig. 6**, Lattice thermal conductivity of silicene, phosphorene, penta-graphene, F-diamane, and $MoS_2$ monolayers by the MTP-based BTE solution and full-DFT counterparts. The DFT-based BTE results are from (a) Gu 2015 [59], Han 2016 [60], Peng 2016 [61] and Liu 2017 [62], (b and c) Zhang 2017 [63], Zheng 2016 [64] and Zhu 2014 [65], (d) Qian 2017 [66], Wang 2016 [67] and Liu 2018 [68], (e) Zhu 2019 [69], (f) Peng 2016 [58], Zulfiqar 2019 [70], Gandi 2015 [71] and Zhao 2018 [72].

Conducted comparisons between the MTP-based BTE solutions and full-DFT counterparts for considered bulk and 2D lattices confirm the remarkable accuracy of the accelerated approach. This is a highly promising finding, as the computational costs are substantially decreased, and at the same time, the sacrificial in accuracy remains marginal. In fact, since the MTP-based approach enables the consideration of larger supercells and cutoff distances in the evaluation of anharmonic interatomic force constants, they can yield more accurate results than simplified DFT-based solutions. By decreasing the symmetry in the atomic lattice and increasing the number of atoms in the primitive unit cell, the number of structures for the force constant calculations increases substantially. These calculations can get exceedingly expensive using the DFT-based approach, whereas with the MTP-based solution, the computational costs always stay negligible. In comparison with the DFT-based solution, the proposed approach is straightforward and does not require the convergence tests with respect to the K-point grid and plane-wave cutoff energy. In the MTP-based approach, independent of lattice complexity, relatively short AIMD trajectories of less than 4000 time steps are required for the training of accurate interatomic potentials. In our earlier study[26], our extensive results for diverse and complex 2D lattices confirm that for the majority of cases the MTPs trained over only 1000 time steps of AIMD simulations at 50 K can very accurately reproduce the phonon dispersion relations and other phononic properties in comparison with DFPT results. However, to improve the accuracy for the evaluation of anharmonic interatomic force constants the inclusion of AIMD trajectories at high temperatures is highly recommended. We remind that even for the case of a highly symmetrical lattice of graphene, the proposed approach is highly efficient from the computational point of view. Nonetheless, the computational efficiency of MTP-based solution will be even more obvious when moving from high- to low-symmetry structures and in particular for 2D lattices that require the consideration of interactions with distant neighbours. Our results in another word suggest that MTP can accurately substitute the DFT method in the evaluation of interatomic force constants. As it has been confirmed in recent studies for the case of graphene [73] and bulk BAs[74], the inclusion of four-phonon scattering can be a critical issue to accurately estimate the thermal conductivity of particular structures. In order to take into count the four-phonon scattering in the solution of thermal conductivity, MTP can offer an outstanding application prospect owing to its inehrent negligible computational cost. We therefore hope that MTP-based approach can serve as a standard and versatile tool to conveniently and accurately examine the lattice thermal conductivity.

## 4. Conclusion

Combination of DFT simulations and BTE solution is currently the most trusted first-principles method to examine the lattice thermal conductivity. The main computational bottleneck of this method is to acquire the anharmonic interatomic force constants using the DFT calculations. In this work, we show that MTPs trained over short ab-initio molecular dynamics trajectories can substantially accelerate the evaluation of anharmonic interatomic force constants. Comparison between the MTP-based BTE solutions and full-DFT counterparts for the thermal conductivity of several bulk and 2D structures confirm the remarkable accuracy of the proposed approach. In order to facilitate the practical employment of the proposed approach in conjunction with the ShengBTE package for the BTE solution, elaborated computational details and all considered examples are included in the data availability section. MTP/ShengBTE approach is therefore expected to serve as a standard tool to conveniently, efficiently, and accurately examine the lattice thermal conductivity, which may otherwise require excessive computational resources with commonly employed full-DFT counterpart.

### Acknowledgment
B.M. and X.Z. appreciate the funding by the Deutsche Forschungsgemeinschaft (DFG, German Research Foundation) under Germany's Excellence Strategy within the Cluster of Excellence PhoenixD (EXC 2122, Project ID 390833453). E.V.P, I.S.N., and A.V.S. were supported by the Russian Science Foundation (Grant No 18-13-00479).

### Data availability
The following data are available to download via http://dx.doi.org/10.17632/fmkvzbk3nt.1 : (1) python scripts developed for the ShengBTE/MLIP interface, (2) a guide for passive training of MTPs using the MLIP package, (3) examples of VASP input scripts for the AIMD simulations, (4) samples of untrained MTPs, (5) and most importantly complete ShengBTE input files for all the considered examples along with the numerical procedure to extract the anharmonic force constants for every example using the trained MTPs.

## 1. Training a moment tensor potential (MTP).

### 1.1 Access to the MLIP package.

MLIP is a software package implementing MTP. It is distributed upon sending a reasonable request to Alexander Shapeev at a.shapeev@skoltech.ru.

### 1.2 Creating training sets.

Training sets are created by running ab-initio molecular dynamics (AIMD) at different temperatures using the *Vienna Ab-initio Simulation Package* (VASP)[1–3]. In the Mendeley dataset, the folder "AIMD-inputs", two samples of VASP input files (namely, POSCAR, POTCAR, INCAR and KPOINTS) for silicene monolayer and bulk silicon are included. After the completion of AIMD simulations, the OUTCAR file can be used to create the training set (train.cfg) with the following command:

```
./mlp convert-cfg OUTCAR train.cfg --input-format=vasp-outcar
```

This converts the configurations to a recognizable file format that is later used for training routine. The training set now contains the correlated configurations and can be reduced (subsampled) using the following command:

```
./mlp subsample train.cfg subsample.cfg 5
```

Here each one out of every 5 snapshots in the original "train.cfg" will be written to "subsample.cfg". The subsampled training sets at different temperatures or structures should then be merged together to create the final training set, which can be achieved using the Linux cat *command*.

### 1.3 Training of MTPs.

Training of MTPs is done by solving the following minimization problem:

$$\sum_{k=1}^{K}\left[w_e\left(E_k^{\text{AIMD}}-E_k^{\text{MTP}}\right)^2 + w_f \sum_i^N |f_{k,i}^{\text{AIMD}}-f_{k,i}^{\text{MTP}}|^2 + w_s \sum_{i,j=1}^{3} |\sigma_{k,ij}^{\text{AIMD}}-\sigma_{k,ij}^{\text{MTP}}|^2\right] \to \min,$$

where $E_k^{\text{AIMD}}$, $f_{k,i}^{\text{AIMD}}$ and $\sigma_{k,ij}^{\text{AIMD}}$ are the energy, atomic forces, and stresses in the training set, respectively, and $E_k^{\text{MTP}}$, $f_{k,i}^{\text{MTP}}$, and $\sigma_{k,ij}^{\text{MTP}}$ are the corresponding values calculated with the MTP, $K$ is the number of the configurations in the training set, $N$ is the number of atoms in a configuration and $w_e$, $w_f$ and $w_s$ are the non-negative weights that express the importance of energies, forces, and stresses in the optimization problem, respectively, which in our study were set to 1, 0.1 and 0.001, respectively. We note that the weights for energy and stress are the default values.

As an example, the training of a MTP can be achieved using the following command:

```
mpirun -n n_cores ./mlp train p.mtp train.cfg --energy-weight=1 --force-weight=0.1 --stress-weight=0.001 --max-iter=3000 --curr-pot-name=p.mtp --trained-pot-name=p.mtp
```

Here "n_cores" is the number of cores used for parallel training of MTP, "p.mtp" is the input/output (curr-pot-name/trained-pot-name) MTP file, "train.cfg" is the training set in the internal *.cfg MLIP format, the option "max-iter" determines the maximum number of iterations in the optimization algorithm. The options "energy-weight", "force-weight", and "stress-weight", respectively, define the $w_e$, $w_f$ and $w_s$ weights explained earlier.

In our work, we conducted the passive training, by parameterizing the MTPs using the subsampled AIMD trajectories. In this approach, from the complete sets of AIMD configurations, only subsamples are selected for the training of first MTPs. Nonetheless, some critical configurations that could result in the improved accuracy of trained MTPs may have been missed in the created subsamples. Therefore, the accuracy of the developed MTP "p.mtp" over current subsampled training set "train.cfg" should once again be checked over the full AIMD configurations "trainF.cfg", and the configurations with high extrapolations grades [4] will be selected, and will written to the file "trainN.cfg", *via the following command:*

```
./mlp  select-add p.mtp train.cfg trainF.cfg trainN.cfg
```

The selected configurations "trainN.cfg" should be added to the original training sets "train.cfg" and the final MTP will developed by retraining of new clean potentials over the updated training set. This way, the efficient use of conducted AIMD simulations will be guaranteed.

### 1.4 Structure of MTPs.

MTP belongs to the family of machine-learning interatomic potentials by which potentials show flexible functional form that allows for systematically increasing the accuracy with an increase in the number of parameters and the size of the training. In the folder "Untrained-MTPs", we included three samples of clean MTPs. Depending on the number of parameters, the appropriate MTP should be chosen. Prior to training, there are some parameters to be adjusted, such as the "species_count", "min_dist" and "max_dist" which, respectively, define the number of elements in the system, minimum atomic distance and cutoff distance of the potential. Like the classical potentials, by increasing the cutoff distance more neighbors will be included in the calculations which accordingly increase the computational costs. The number of parameters in a MTP can be calculated via:

$$\text{species\_count}^2 \cdot \text{radial\_basis\_size} \cdot \text{radial\_funcs\_count} + \text{alpha\_scalar\_moments} + 1$$

Note that "radial_funcs_count" and "alpha_scalar_moments" are the fixed features of a particular MTP and only "radial_basis_size" can be manually changed to adjust the number of constants.

### 2. Evaluation of phononic properties using the MTPs.

In our previous work [5], we included the full details and numerous examples for the evaluation of phononic properties using the MTP and PHONOPY [6] package in a public Mendeley dataset, please refer to: http://dx.doi.org/10.17632/7ppcf7cs27.1

### 3. MTP/ShengBTE interface.

ShengBTE [7] is a package for computing the lattice thermal conductivity on the basis of a full iterative solution to the Boltzmann transport equation. Its main inputs are sets of second- and third-order interatomic force constants and a `CONTROL` file for the adjustment of computational details. In this work, the calculation of anharmonic interatomic force constants is substantially accelerated by substituting DFT simulations with the MTP-based solution. For the calculation of anharmonic interatomic force constants, ShengBTE [7] provides a script, *"thirdorder.py"*, implementing a real-space supercell approach to anharmonic IFC calculations. In this approach, according to the defined supercell size and cutoff distance, the input geometries for the force constant calculations will be generated. For compatibility with "`cfg`"-file format, the "`thirdorder_mtp.py`" is developed using the original "`thirdorder_vasp.py`". Moreover, we developed an additional script "`fake_vasp_calcs.py`", which uses the MTP-based calculated forces and artificially create the VASP output files of "`vasprun.xml`". This approach provides the possibility of direct comparison of forces by MTP and VASP. These developed two python scripts are included the folder "`MTP_ShengBTE_py`".

In the folder "`Examples`", complete input files are included for every structure. In this case the subfolder "`ShengBTE-inputs`" includes the complete input files for the ShengBTE solution (namely: `CONTROL`, `FORCE_CONSTANTS_2ND` and `FORCE_CONSTANTS_3RD`). Using the data provided in the subfolder called "`Anharmonic-MTP`", the anharmonic interatomic force constants can be obtained using the trained MTPs "`p.mtp`". MTP/ShengBTE interface follows the same routine as that of the VASP/ShengBTE, explained in the ShengBTE documentation. To facilitate the practical usage, for every example we included a shell script, named "`getFC.sh`". In the aforementioned script, "`supcell`" and "`Cutoff`" are respectively, the supercell size and cutoff neighbour for the evaluation of anharmonic force constants on the basis of primitive unitcell "`POSCAR`". Please note that "`POSCAR`", "`p.mtp`", "`mlip.ini`" and related python scripts should be located in this folder for complete calculations.